\documentclass[preprint,showpacs,preprintnumbers,amssymb,aps]{revtex4}

\usepackage{graphicx}
\usepackage{dcolumn}
\usepackage{amsmath}
\usepackage{epsfig}
\usepackage{subfigure}
\usepackage{array}
\usepackage{epstopdf} 
\usepackage{xcolor}

\graphicspath{{figures/}}

\begin{document}

\normalsize

\title{\boldmath  Next-To-Leading Order Differential Cross-Sections for $J/\psi$, $\psi(2S)$ and $\Upsilon$ Production
in Proton-Proton Collisions at a Fixed-Target Experiment using the LHC Beams (AFTER@LHC)}

\author{
\small Yu Feng and Jian-Xiong Wang \\
 {\it Institute of High Energy Physics, Chinese Academy of Sciences, P.O.Box 918(4), Beijing 100049, China.
 }}

\begin{abstract}

  Using  nonrelativistic QCD (NRQCD) factorization, we calculate the yields for $J/\psi$, $\psi(2S)$ and $\Upsilon(1S)$ hadroproduction at $\sqrt{s}=$ 72 GeV and 115 GeV including the next-to-leading order QCD corrections.
  Both these center-of-mass energies correspond to those obtained with 7~TeV and 2.76~TeV nucleon
  beam impinging a fixed target.
  We study the cross section integrated in $p_t$ as a function of the rapidity as well as the $p_t$ differential cross section in the central rapidity region.
  Using different NLO fit results of the  NRQCD long-distance matrix elements,
  we evaluate a theoretical uncertainty which is certainly much larger than the projected experimental uncertainties with the expected 20 fb$^{-1}$ to be collected per year with AFTER@LHC.

\end{abstract}


\maketitle

\section{Introduction}

Non-relativistic quantum chromodynamics (NRQCD)~\cite{Bodwin:1994jh} is
the most systematic factorization scheme to describe the decay and production of heavy quarkonia.
It allows one to organize the theoretical calculations as double expansions in both
the coupling constant $\alpha_s$ and the heavy-quark relative velocity $v$.
In the past few years, significant progress has been made in next-to-leading order (NLO) QCD calculations based on NRQCD.
Calculations and fits of NRQCD long-distance matrix elements (LDMEs) for
both the $J/\psi$ yield and polarization in hadroproduction have been
carried out~\cite{Ma:2010yw,Butenschoen:2010rq,Butenschoen:2012px,Chao:2012iv,Gong:2012ug} as well as
for $\Upsilon$ hadroproduction~\cite{Wang:2012is,Gong:2013qka}.
Using these LMDEs, one can in principle predict the transverse momentum $p_t$ differential cross section at any energies.
In addition, in a recent study~\cite{Feng:2015cba}, we have discussed the implication of these fits on the energy dependence of the cross sections integrated in $p_t$.

In this paper, we predict these differential cross sections for the kinematics
of a fixed-target experiment using the LHC beams (AFTER@LHC)~\cite{Brodsky:2012vg}.
In practice, 7 TeV protons on targets yield to a c.m.s energy close to
115 GeV  and 72 GeV for 2.76 TeV nucleons (as in the case of a Pb beam).
This corresponds to a range very seldom explored so far, significantly
higher than that at CERN-SPS and not far from BNL-RHIC.
With the typical luminosity of the fixed-target mode, which allows for
yearly luminosities as large as 20 fb$^{-1}$, AFTER@LHC is expected to
be a quarkonium and heavy-flavor
observatory~\cite{Brodsky:2012vg,Lansberg:2012kf}. In general, the
opportunities of a fixed-target experiment using the LHC beam for spin
and heavy-ion physics are discussed
in~\cite{Brodsky:2012vg,Massacrier:2015nsm,Lansberg:2014myg,Rakotozafindrabe:2012ei}.
In this work, we confirm that charmonium yields can easily reach $10^9$
per year and $10^6$ for bottomonia.

\section{Next-to-leading order calculation}

Following the NRQCD factorization formalism~\cite{Bodwin:1994jh}, the cross section for quarkonium hadroproduction $H$
can be expressed as
\begin{eqnarray}
  d\sigma[pp\rightarrow H + X]=\sum_{i,j,n}\int dx_1 dx_2 G^i_p G^j_p
  d\hat{\sigma}[ij\rightarrow (Q\overline{Q})_n X]\langle{\cal O}^{H}_n \rangle
\end{eqnarray}
where $p$ is either a proton or an antiproton, $G^{i(j)}_p$ is the parton distribution function (PDF) of $p$,
the indices $i,j$ runs over all possible partonic species, and $n$ denotes the color, spin and angular momentum
states of the intermediate $Q\overline{Q}$ pair.
For $\psi$ and $\Upsilon$, namely the $^3S_1$ quarkonium sates, their leading CO states of relative order $\mathcal{O}(v^4)$ are
$^1S_0^{[8]}$, $^3S_1^{[8]}$, $^3P_J^{[8]}$.
Along with the CS transition $^3S_1^{[1]}$, we call the total CS + CO contributions as direct production.
The short-distance coefficient (SDC) $d\hat{\sigma}$ will be calculated perturbatively,
while the long-distance matrix elements (LDMEs) $\langle{\cal O}^{H}_n \rangle$ are governed by nonperturbative QCD effects.

Now let us take a look at the parton level processes related in this work.
As we know that, for hadroproduction, the CO contributions appear at $\alpha_s^2$ ~\cite{Cho:1995ce}, their Born contributions are
\begin{eqnarray}\label{equ:lo-sigma}
\nonumber  &&q+\overline{q} \rightarrow Q\overline{Q} [^3S_1^{[8]}], \\
  &&g+g \rightarrow Q\overline{Q} [^1S_0^{[8]}, ^3P_{J=0,2}^{[8]}],
\end{eqnarray}
where $q(\overline{q})$ denotes the light quarks (untiquarks).

Up to $\alpha_s^3$, QCD corrections include real and virtual parts. One inevitably encounters ultra-violet (UV),
infra-red(IR) and Coulomb divergences when dealing the virtual corrections.
UV divergences from self-energy and triangle diagrams are canceled upon the renormalization procedure.
For the real emission corrections, three kinds of processes are contained
\begin{eqnarray}\label{equ:nlo-sigma}
\nonumber  &&g+g \rightarrow Q\overline{Q} [^3S_1^{[1]}, ^1S_0^{[8]},^3S_1^{[8]}, ^3P_{J=0,2}^{[8]}]+g , \\
  && g+q(\overline{q}) \rightarrow Q\overline{Q}[^1S_0^{[8]},^3S_1^{[8]}, ^3P_{J=0,2}^{[8]}] + q(\overline{q}) , \\
\nonumber  &&q+\overline{q} \rightarrow Q\overline{Q} [^1S_0^{[8]}, ^3S_1^{[8]}, ^3P_{J=0,1,2}^{[8]} ] +g.
\end{eqnarray}
some of which involve IR singularities in phase-space integration and
we adopt the two-cutoff phase space slicing method~\cite{Harris:2001sx} to isolate these singularities
by introducing two small cutoffs, $\delta_s$ and $\delta_c$. For technique details,
we refer readers to Ref.~\cite{Gong:2008hk,Gong:2008ft}.

One has to note that in Eq.(\ref{equ:nlo-sigma}), the $^3S_1^{[1]}$ production in $gg$ fusion is not really correction,
strictly speaking, it is only the Born order contribution for hadroproduction with a jet.
In fact, all the real emission processes in Eq.(\ref{equ:nlo-sigma}) will be taken as Born-order contributions
of quarkonium - jet production.
Then one can discuss the $p_t$ dependent differential cross section and,
the QCD NLO corrections in this case are up to $\alpha_s^4$, which involves real emission processes
\begin{eqnarray}\label{equ:nlo-dpt}
\nonumber  && g+g \rightarrow (Q\overline{Q})_n +g + g , ~~~g+g \rightarrow (Q\overline{Q})_n+q+\overline{q} , \\
\nonumber  && g+q(\overline{q}) \rightarrow (Q\overline{Q})_n +g+ q(\overline{q}) , ~~~q+\overline{q} \rightarrow (Q\overline{Q})_n + g + g, \\
\nonumber  && q+\overline{q} \rightarrow (Q\overline{Q})_n +q+\overline{q}, ~~~q+\overline{q} \rightarrow (Q\overline{Q})_n +q'+\overline{q}', \\
  && q+ q \rightarrow (Q\overline{Q})_n + q + q, ~~~q+ q' \rightarrow (Q\overline{Q})_n + q + q' .
\end{eqnarray}
where $q$, $q'$ denote light quarks with different flavors and $(Q\overline{Q})_n$ can be either
$^3S_1^{[1]}$, $^1S_0^{[8]}$, $^3S_1^{[8]}$, or $^3P_J^{[8]}$.
One can find the detailed descriptions at this order in Ref.~\cite{Gong:2008ft,Gong:2010bk} and some examples
~\cite{Ma:2010yw,Gong:2012ug,Butenschoen:2010rq,Wang:2012is,Gong:2013qka}.

All of these calculations are made with the newly-updated Feynman Diagram Calculation package~\cite{Wang:2004du}.

\section{Constrains on the LDMEs}
The color-singlet (CS) LDMEs are estimated from wave functions at the origin by
$\langle{\cal O}^{H}(^{3}S^{[1]}_{1})\rangle = \frac{3N_c}{2\pi}|R_{H}(0)|^{2}$,
where the wave functions are obtained via potential model calculation~\cite{Eichten:1995ch},
which gives $|R_{J/\psi}(0)|^{2}=0.81$ GeV$^3$, $|R_{\psi(2S)}(0)|^{2}=0.53$ GeV$^3$, $|R_{\Upsilon(1S)}(0)|^{2}=6.5$ GeV$^3$.
We note this part as CSM results when performed separately in the following context.

For the color-octet (CO) LDMEs, they can only be extracted from data.
Due to the improvements of NLO calculation, groups of LDMEs based on NLO corrections
are obtained by different fitting schemes.
Some of them are used in this work as the theoretical uncertainty
and we will give a brief discussion on these CO LDMEs below.

In the $J/\psi$ case, seven groups of LDMEs~\cite{Butenschoen:2011yh,Chao:2012iv,Ma:2010yw,
Gong:2012ug,Zhang:2014ybe,Han:2014jya,Bodwin:2014gia} are collected in Table.~\ref{tab:ldmes-jpsi}.
They are extracted by fitting the data of hadroproduction yield~\cite{Ma:2010yw}, or combined with
polarization~\cite{Chao:2012iv, Gong:2012ug} on $pp$ collisions.
The first one~\cite{Butenschoen:2011yh} was based on a
wider set of data including $ep$ and $\gamma\gamma$ system with $p_t > 1$ GeV.
In Ref.~\cite{Chao:2012iv, Gong:2012ug}, the data with $p_t < 7$ GeV are excluded in their fit.
The fit in Ref.~\cite{Zhang:2014ybe,Han:2014jya} took the $\eta_c$ measurement ($p_t \ge 6$ GeV) into consideration.
Only one of them is used~\cite{Han:2014jya} since their results are almost the same.
The last one incorporates the leading-power fragmentation corrections together with
the QCD NLO corrections, which results in a different SDC and may bring different LDMEs.
In Ref.~\cite{Ma:2010yw}, Ma $et~al.$ fit the data with $p_t > 7$ GeV by two linear combinations of LDMEs:
\begin{eqnarray}
\nonumber M^{J/\psi}_{0,r_{0}}=\langle{\cal O}^{J/\psi}(^{1}S_{0}^{[8]})\rangle + \frac{r_{0}}{m^2_c}\langle{\cal O}^{J/\psi}(^3P_0^{[8]})\rangle \\
M^{J/\psi}_{1,r_{1}}=\langle{\cal O}^{J/\psi}(^{3}S_{1}^{[8]})\rangle + \frac{r_{1}}{m^2_c}\langle{\cal O}^{J/\psi}(^3P_0^{[8]})\rangle
\label{equ:ma}
\end{eqnarray}
where we extract the value of LDMEs by limiting $\langle{\cal O}^{J/\psi}(^{1}S_{0}^{[8]})\rangle$
and $\langle{\cal O}^{J/\psi}(^{3}S_{1}^{[8]})\rangle$ to be positive to get a loose constraint on
the $\langle{\cal O}^{J/\psi}(^3P_0^{[8]})\rangle$ range,
from which we choose the middle value to obtain the three LDMEs (\textit{Ma(2011)} in Table.~\ref{tab:ldmes-jpsi}).

\begin{table}[!ht]
  \begin{center}
  \caption{ \label{tab:ldmes-jpsi} The values of LDMEs for $J/\psi$ hadroproduction (in units of GeV$^3$). }
  \footnotesize
  \begin{tabular*}{155mm}{@{\extracolsep{\fill}}ccccc}
  \hline\hline
   Ref. & $\langle{\cal O}^{J/\psi}(^{3}S^{[1]}_{1})\rangle$ ~&~ $\langle{\cal O}^{J/\psi}(^{1}S^{[8]}_{0})\rangle$
  ~&~ $\langle{\cal O}^{J/\psi}(^{3}S^{[8]}_{1})\rangle$ ~&~ $\langle{\cal O}^{J/\psi}(^{3}P^{[8]}_{0})\rangle/m_Q^2$  \\
  \hline
  Butenschoen(2011)~\cite{Butenschoen:2011yh} ~&~ 1.32 ~&~ $3.0\times 10^{-2}$ ~&~ $1.7\times 10^{-3}$ ~&~ $-4.0 \times 10^{-3}$  \\
  Chao(2012)~\cite{Chao:2012iv}   ~&~  1.16 ~&~ $8.9\times 10^{-2}$ ~&~ $ 3.0\times 10^{-3}$  ~&~ $ 5.6 \times 10^{-3}$  \\
  Ma(2011)~\cite{Ma:2010yw}   ~&~  1.16 ~&~ $3.9\times 10^{-2}$ ~&~ $ 5.6\times 10^{-3}$  ~&~ $ 8.9 \times 10^{-3}$  \\
  Gong(2013)~\cite{Gong:2012ug} ~&~  1.16 ~&~ $9.7\times 10^{-2}$ ~&~ $-4.6\times 10^{-3}$  ~&~ $-9.5\times 10^{-3}$  \\
  Zhang(2015)~\cite{Zhang:2014ybe} ~&~ $0.24\sim0.90$  ~&~ $(0.4\sim1.1)\times10^{-2}$ ~&~ $1.0\times 10^{-2}$ ~&~ $1.7\times10^{-2}$ \\
  Han(2015)~\cite{Han:2014jya} ~&~ 1.16 ~&~ $0.7\times 10^{-2}$ ~&~ $1.0\times 10^{-2}$ ~&~  $1.7 \times 10^{-2}$ \\
  Bodwin(2014)~\cite{Bodwin:2014gia} ~&~ 0 ~&~ $9.9\times 10^{-2}$ ~&~ $1.1 \times 10^{-2}$ ~&~  $4.9\times 10^{-3}$ \\
  \hline\hline
  \end{tabular*}
  \end{center}
  \end{table}

\begin{table}[!b]
  \begin{center}
  \caption{ \label{tab:ldmes-psi-upsi} The values of LDMEs for $\psi(2S)$ and $\Upsilon(1S)$ hadroproduction (in units of GeV$^3$). }
  \footnotesize
  \begin{tabular*}{155mm}{@{\extracolsep{\fill}}cccccc}
  \hline\hline
  H & Ref. & $\langle{\cal O}^{H}(^{3}S^{[1]}_{1})\rangle$ ~&~ $\langle{\cal O}^{H}(^{1}S^{[8]}_{0})\rangle$ ~&~$\langle{\cal O}^{H}(^{3}S^{[8]}_{1})\rangle$ ~&~
  $\langle{\cal O}^{H}(^{3}P^{[8]}_{0})\rangle/m_Q^2$  \\
  \hline
  $\psi(2S)$ & Gong(2013)~\cite{Gong:2012ug} ~&~  0.76  ~&~$-1.2\times 10^{-4}$ & $3.4\times 10^{-3}$ & $4.2\times 10^{-3}$   \\
     &  Ma(2011)~\cite{Ma:2010yw}     ~&~  0.76  ~&~$ 1.4\times 10^{-2}$ & $2.0\times 10^{-3}$ & $1.6\times 10^{-3}$   \\
  $\Upsilon(1S)$ & Gong(2014)~\cite{Gong:2013qka} ~&~    9.28  ~&~ $11.2\times10^{-2}$ ~&~ $-4.1\times 10^{-3}$ ~&~ $-6.7\times 10^{-3}$ \\
  &  Han(2014)~\cite{Han:2014kxa} ~&~    9.28  ~&~ $3.5\times10^{-3}$ ~&~ $5.8\times 10^{-2}$ ~&~ $3.6\times 10^{-2}$ \\
   & Feng(2015)~\cite{Feng:2015wka} ~&~    9.28  ~&~ $13.6\times10^{-2}$ ~&~ $ 6.1\times 10^{-3}$ ~&~ $ -9.3\times 10^{-3}$ \\
  \hline\hline
  \end{tabular*}
  \end{center}
  \end{table}

As regards the $\psi(2S)$, only two NLO analyses results in Ref.~\cite{Ma:2010yw,Gong:2012ug} are used,
both of which excluded the data with $p_t < 7$ GeV in their fit.
To extract the LDMEs value from the fitting results of Ma $et~al.$, the same method is used as for the $J/\psi$.
For $\Upsilon(1S)$, we use three groups of LDMEs~\cite{Gong:2013qka,Han:2014kxa,Feng:2015wka}.
Both of them have separated the direct production and the feed-down contributions exactly.
In the fit of Ref.~\cite{Han:2014kxa}, only the data in $p_t > 15$ GeV region are used,
while in Ref.~\cite{Gong:2013qka,Feng:2015wka} the region is $p_t > 8$ GeV.
They all describe the high $p_t$ yield data at Tevatron and LHC very well.
We gather the LDMEs of $\psi(2S)$ and $\Upsilon(1S)$ in Table.~\ref{tab:ldmes-psi-upsi}.

\section{Numerical results}

The differential cross sections with rapidity distribution and transverse momentum distribution are considered in the calculation.
In both cases, the CTEQ6M parton distribution functions~\cite{Pumplin:2002vw} and corresponding
two-loop QCD coupling constants $\alpha_s$ are used. The charm quark mass is set to be $m_c=$ 1.5 GeV, while for
bottom quark it is $m_b=$ 4.75 GeV. The renormalization and factorization scales are chosen as $\mu_r=\mu_f=2m_Q$
for rapidity distribution plots, while for the plots of transverse momentum distribution they are $\mu_r=\mu_f=\mu_T$,
with $\mu_T=\sqrt{(2m_Q)^2+p_t^2}$. NRQCD scale is chosen as $\mu_{\Lambda}=m_Q$.
It is important to note that different choices of these scales may be adopted for the CO LDMEs we used from
different groups, which can bring some uncertainties in our prediction.
The uncertainties from scales and quark masses are also considered for cross sections with rapidity distribution, where
scale dependence is estimated by varying $\mu_r$, $\mu_f$, by a factor of 1/2 and 2 with respect to their central values
and quark masses varying 0.1 GeV up and down for $J/\psi$, as well as 0.25 GeV for $\Upsilon$.
Branching ratios are taken from PDG~\cite{Agashe:2014kda}, which give $\mathcal{B}[J/\psi\rightarrow\mu\mu]$ = 0.0596,
$\mathcal{B}[\psi(2S)\rightarrow\mu\mu]$ = 0.0079 and $\mathcal{B}[\Upsilon(1S) \rightarrow\mu\mu]$ = 0.0248, respectively.
The two phase space cutoffs $\delta_s=10^3$ and $\delta_c=\delta_s/50$ are chosen
and the insensitivity of the results on different choices for these cutoffs has been checked.

\subsection{$d\sigma/dy$ up to $\alpha_s^3$}

\begin{figure}[!ht]
  \centering
  \includegraphics[width=8cm]{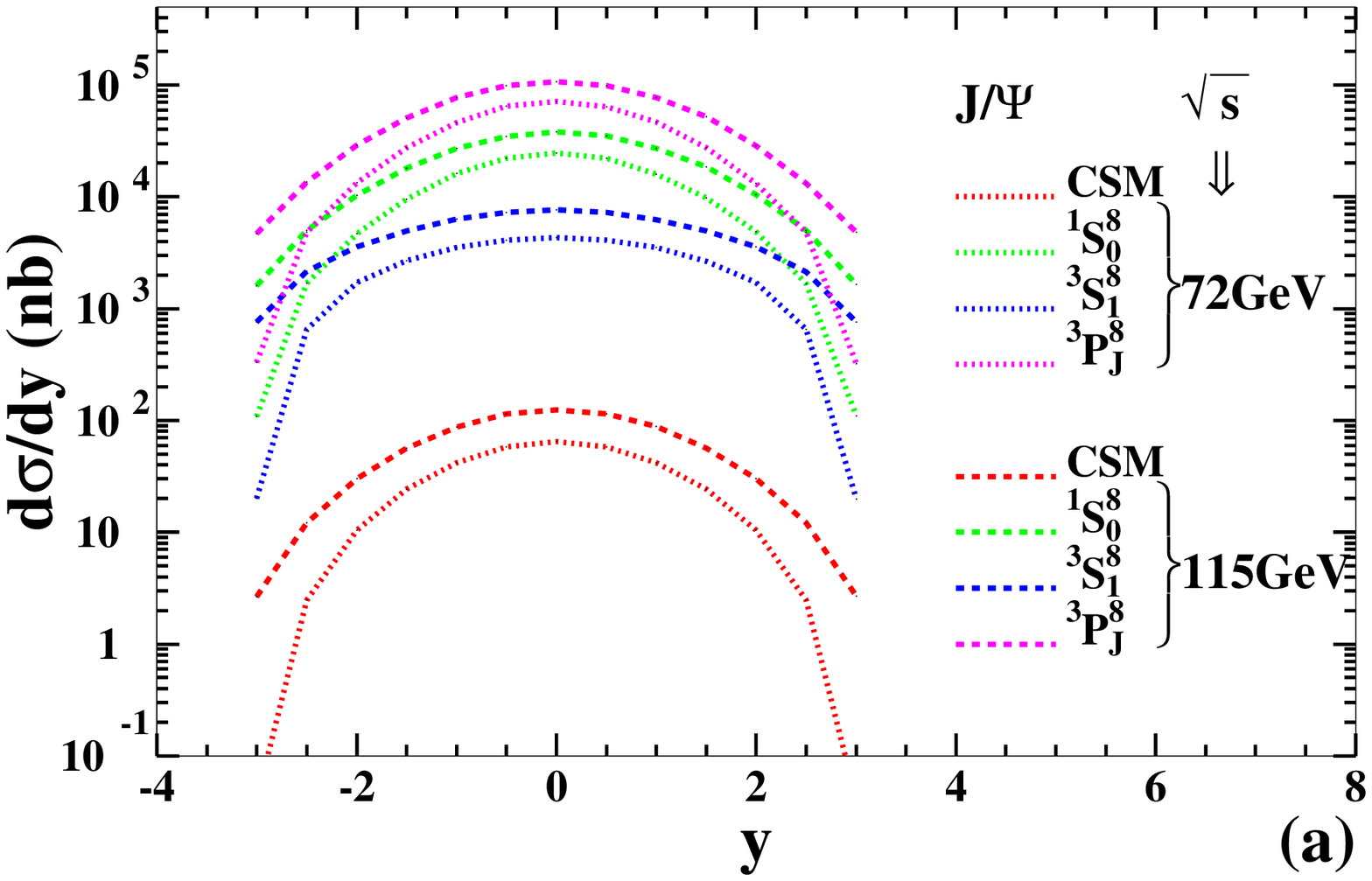}  \includegraphics[width=8cm]{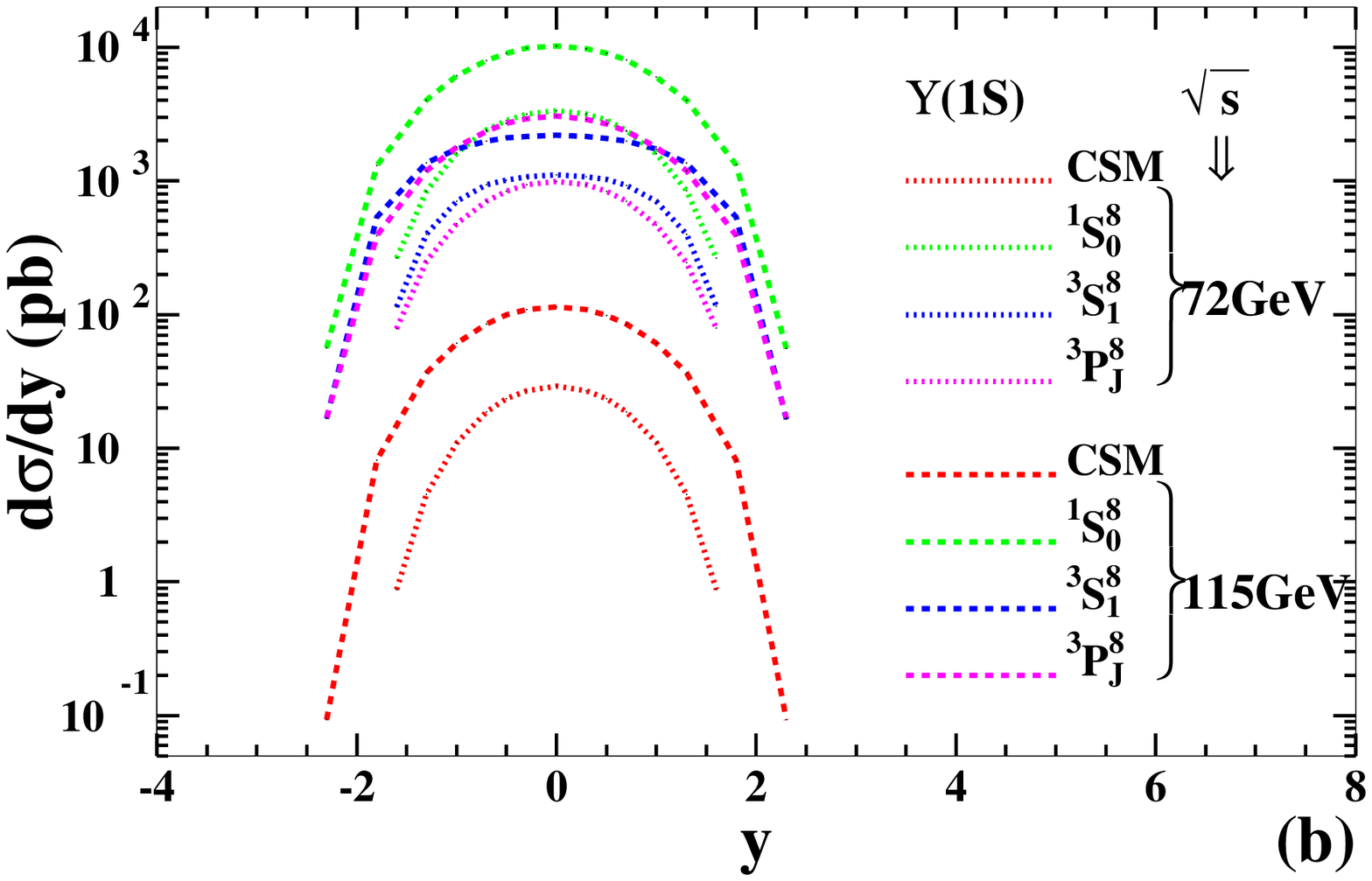}\\
  \caption{Branching contributions of the cross section for direct $J/\psi$ (left) and $\Upsilon(1S)$ (right)
  hadroproduction at the c.m.s energy 72 GeV (dot lines) and 115 GeV (dashed lines), respectively.
  The CO LDMEs for all the channels are set to unity.}
  \label{fig:y-chan}
\end{figure}

\begin{figure}[!ht]
  \centering
  \includegraphics[width=8cm]{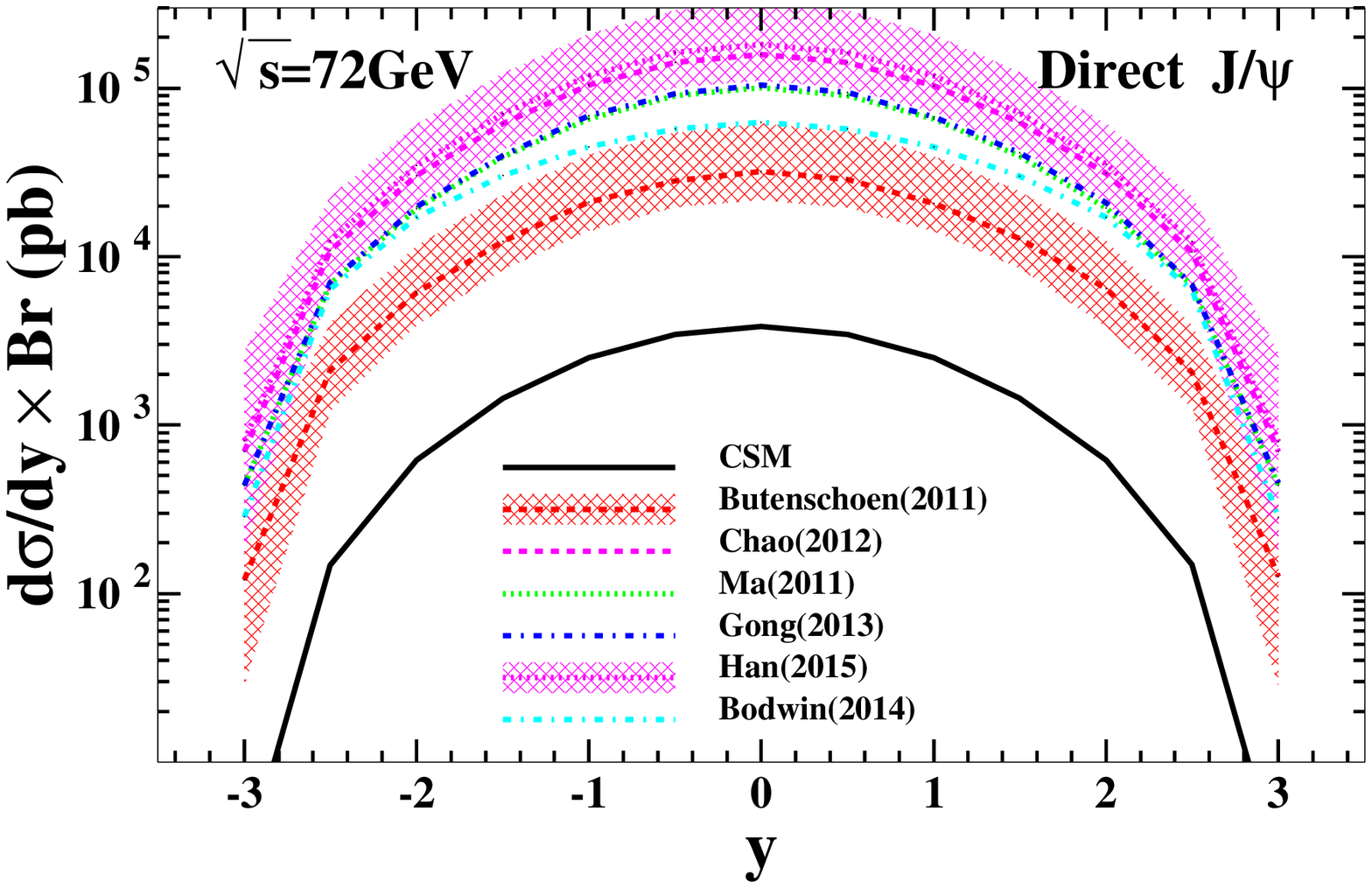}  \includegraphics[width=8cm]{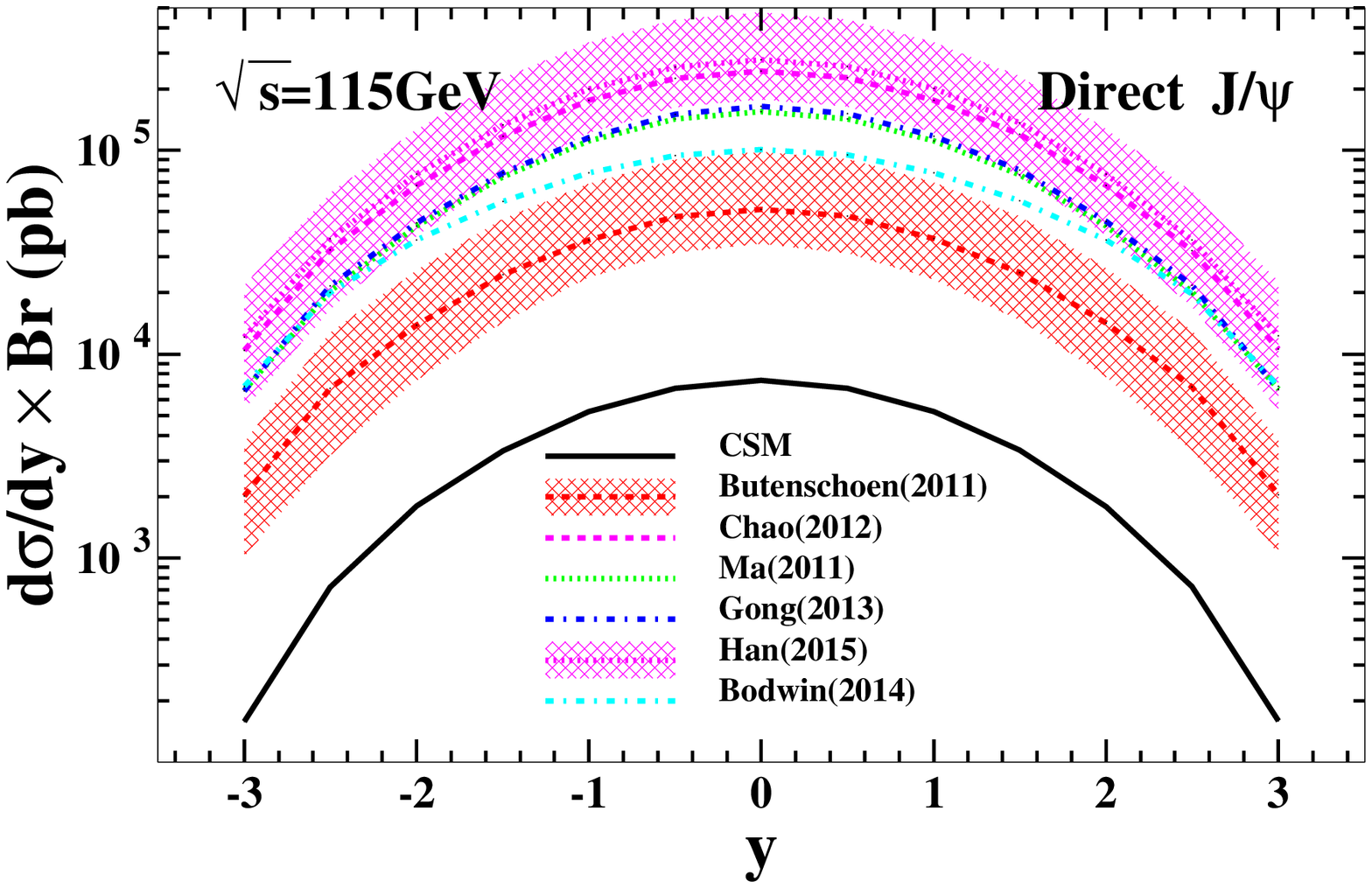} \\
  \includegraphics[width=8cm]{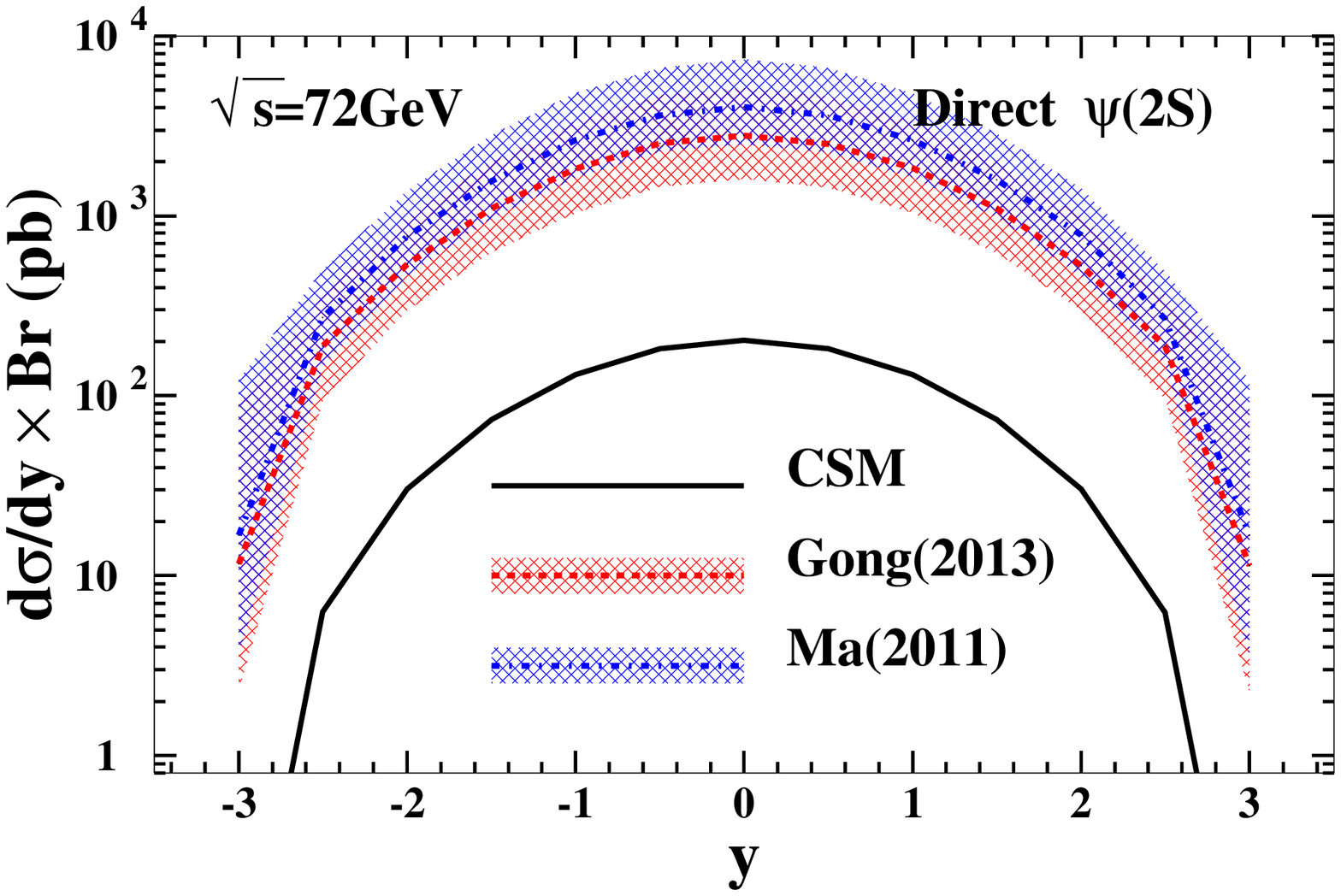}   \includegraphics[width=8cm]{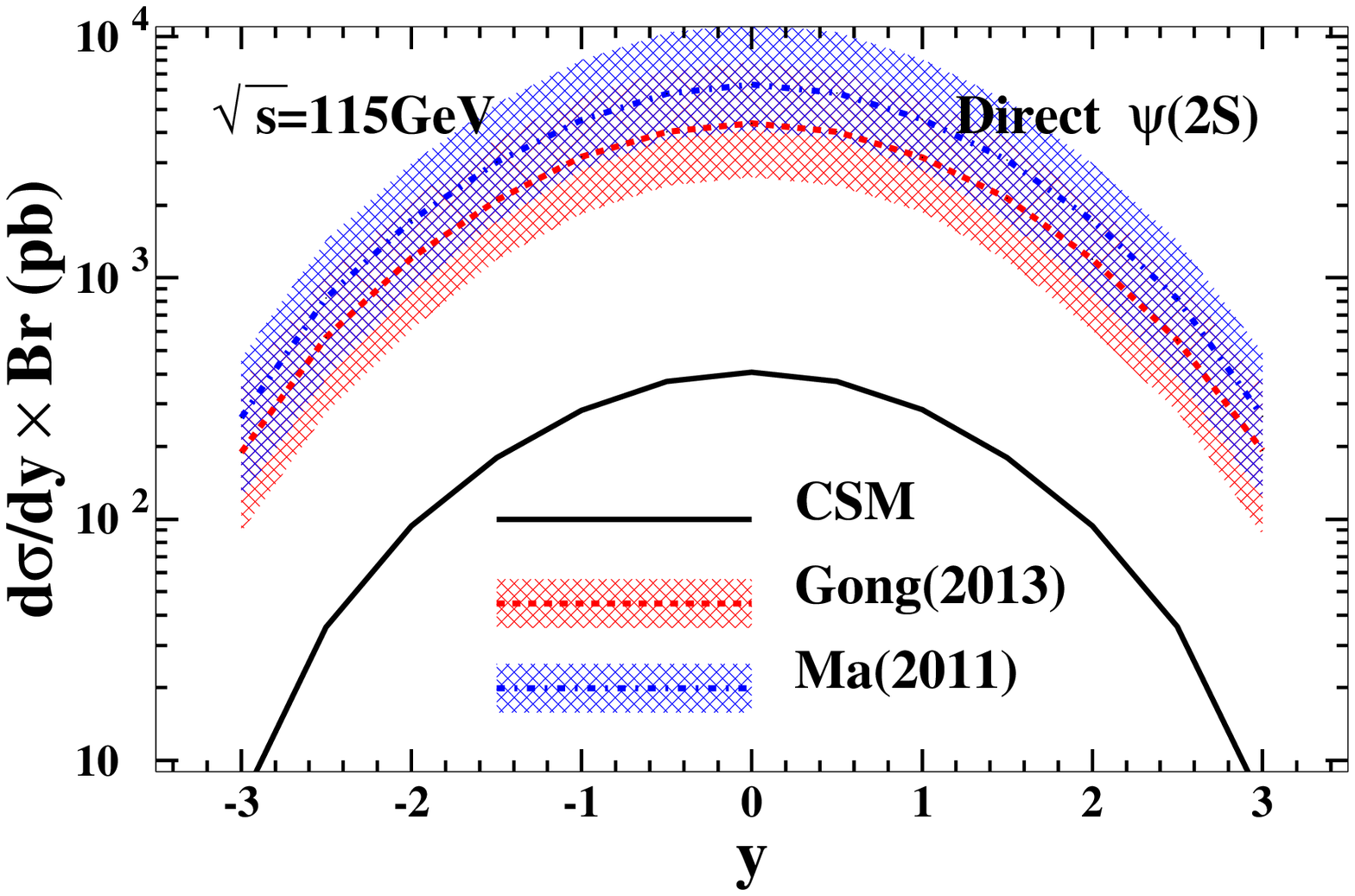} \\
  \includegraphics[width=8cm]{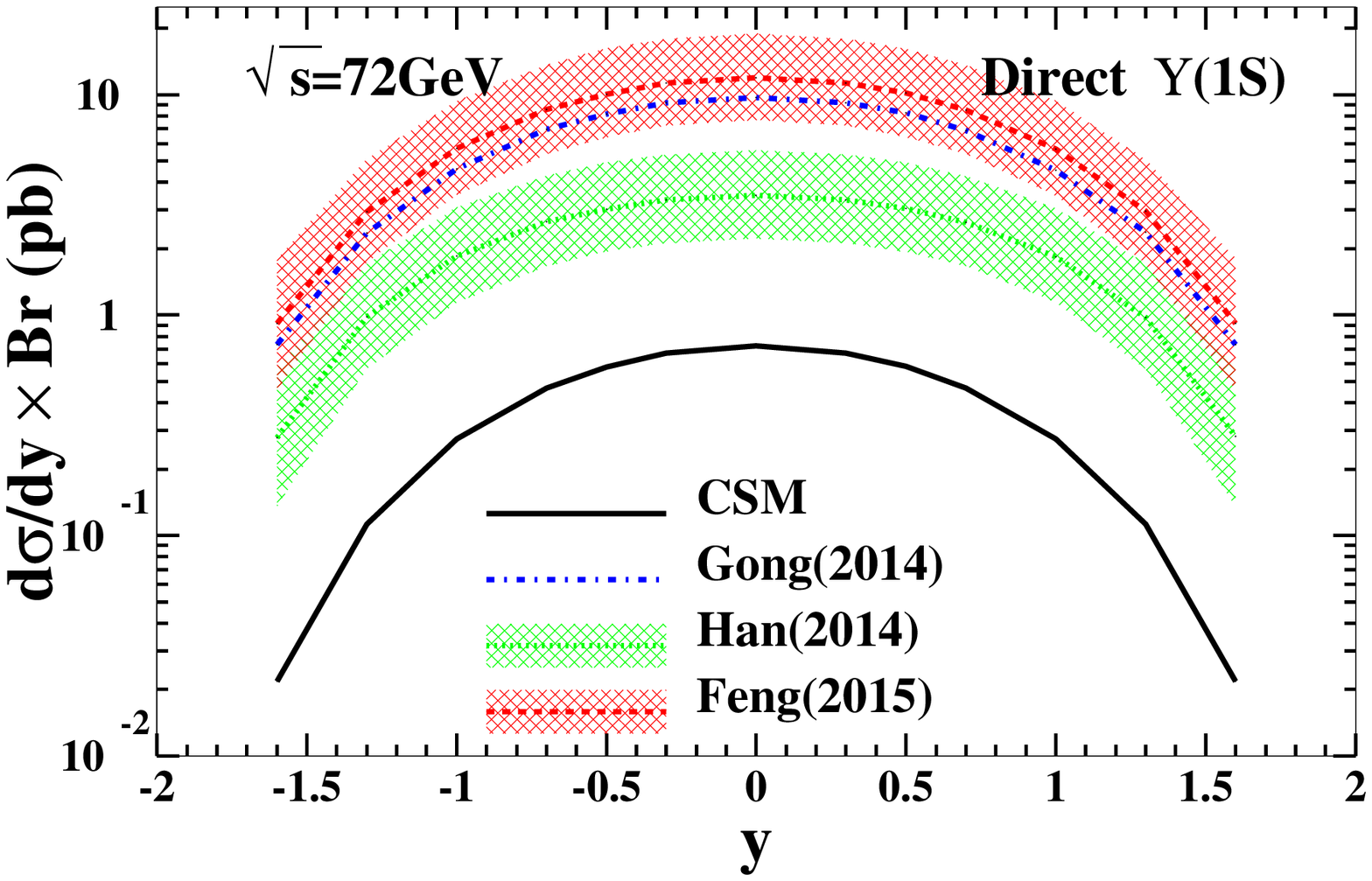}  \includegraphics[width=8cm]{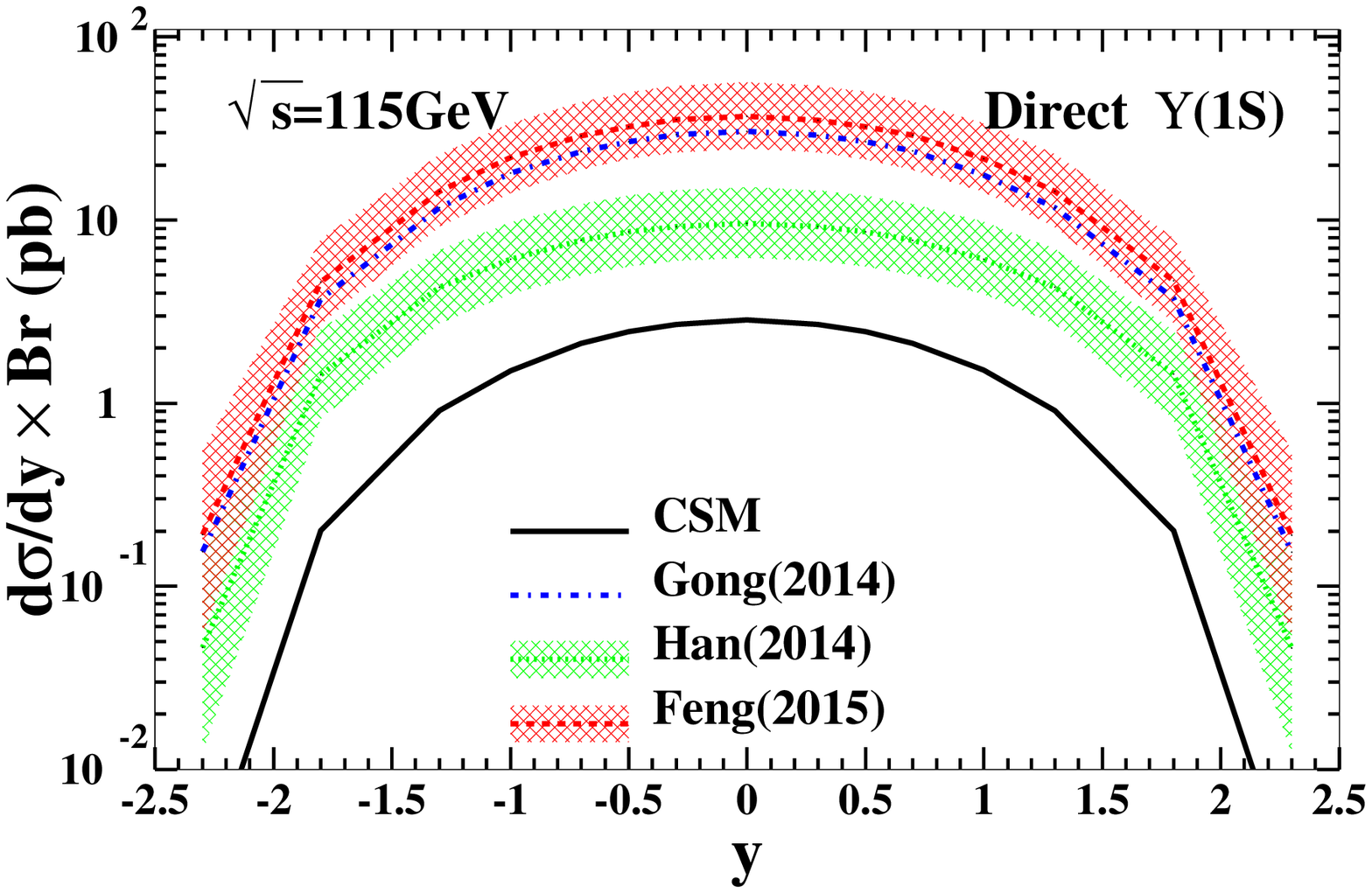} \\
  \caption{Rapidity distribution of differential cross section for direct $J/\psi$ (top), $\psi(2S)$ (middle)
  and $\Upsilon(1S)$ (bottom) hadroproduction at the center of mass energy $\sqrt{s}=72$ GeV
  and $\sqrt{s}=115$ GeV, respectively. The lines are the uncertainty from LDMEs values
  while the color areas are scales and masses uncertainties. }
  \label{fig:y}
\end{figure}

We study the $p_t$ integrated cross section (where the whole $p_t$ region are integrated)
as a function of rapidity in this subsection.
The QCD NLO corrections are up to $\alpha_s^3$ here.
In Fig.~\ref{fig:y-chan} and Fig.~\ref{fig:y}, we perform the rapidity distribution of direct $J/\psi$,
$\psi(2S)$ and $\Upsilon(1S)$ production cross section at center of mass energy $\sqrt{s}=72$ GeV and 115 GeV, respectively.
We first discuss the branching contributions shown in Fig.~\ref{fig:y-chan}, where the CO LDMEs are set to unity for all three production channels.
For $\psi(2S)$, the CSM is different from $J/\psi$ only by a factor, we therefore do not perform it separately.
Obviously, the CSM results (red lines) for both $J/\psi$ and $\Upsilon(1S)$ is small compared with
the CO channels. The dominant CO channel for $J/\psi$ is $^3P_J^{[8]}$ transition,
while for $\Upsilon(1S)$ it is $^1S_0^{[8]}$. Besides, the branching contributions for $J/\psi$ have visible hierarchy,
but for $\Upsilon(1S)$, little difference between $^3S_1^{[8]}$ and $^3P_J^{[8]}$ contributions.

Adopting the LDMEs in Table.~\ref{tab:ldmes-jpsi} and ~\ref{tab:ldmes-psi-upsi}, we present the rapidity distribution
of cross section for various cases in Fig.~\ref{fig:y}.
The lines are the central values with different groups of LDMEs, while the colored areas are the uncertainties
from scales and quark masses. Only the boundary lines are shown with scales and mass uncertainties.
For the $J/\psi$, six groups of NRQCD results are shown as a band,
the boundaries of which has a distance within factor 10.
The values of the cross sections are roughly in the region of $10^4\sim 10^5$ pb.
The CSM results lower than the band, again by a factor 10.
Without a surprise, the CSM seems to be negligible for total NRQCD results.
However, it may not be the case. In fact, as we have discussed in Ref.~\cite{Feng:2015cba},
the LO CSM contribution explains the data very well from the RHIC to LHC energies, while the CO LDMEs extracted from
$p_t$-differential NLO correction would lead to the $p_t$-integrated cross section overshooting the data.
Only the fits from Butenschoen $et~al.$~\cite{Butenschoen:2011yh} that including rather low $p_t$ data provides an
acceptable description of the pt-integrated cross section.
Based on these discussion, most of the predictions in Fig.~\ref{fig:y} might overshoot
the data and CSM may underestimate the measurements below RHIC energy.
For various groups of the LDMEs, they are fitted with large $p_t$ data,
while in our calculation the whole $p_t$ region are integrated.
We suppose the one of Butenschoen $et~al.$~\cite{Butenschoen:2011yh}, namely the lower boundary of the band
(red dashed line) would gives a best prediction for $J/\psi$,
though their LDMEs will meet difficulty when describing the polarization data.

As regards the $\psi(2S)$, two groups of LDMEs lead to a consistent predictions which give the cross section around
$10^3$ pb at both $\sqrt{s}=$ 72 GeV and 115 GeV.
With the uncertainties of scales and quark masses, the cross sections reach $10^4$ pb in the central rapidity region.
Nevertheless, these results overestimated the data as discussed in
Ref.~\cite{Feng:2015cba}.

In the $\Upsilon(1S)$ case, two curves are close and the left one is slightly departure. Yet, their difference is only in
pb units. We ever performed a quite good prediction for $\Upsilon(1S)$ at RHIC energies and below~\cite{Feng:2015cba},
which includes the energies we considered here.


\subsection{$d\sigma/dp_t$ up to $\alpha_s^4$}
\begin{figure}
  \centering
  \includegraphics[width=8cm]{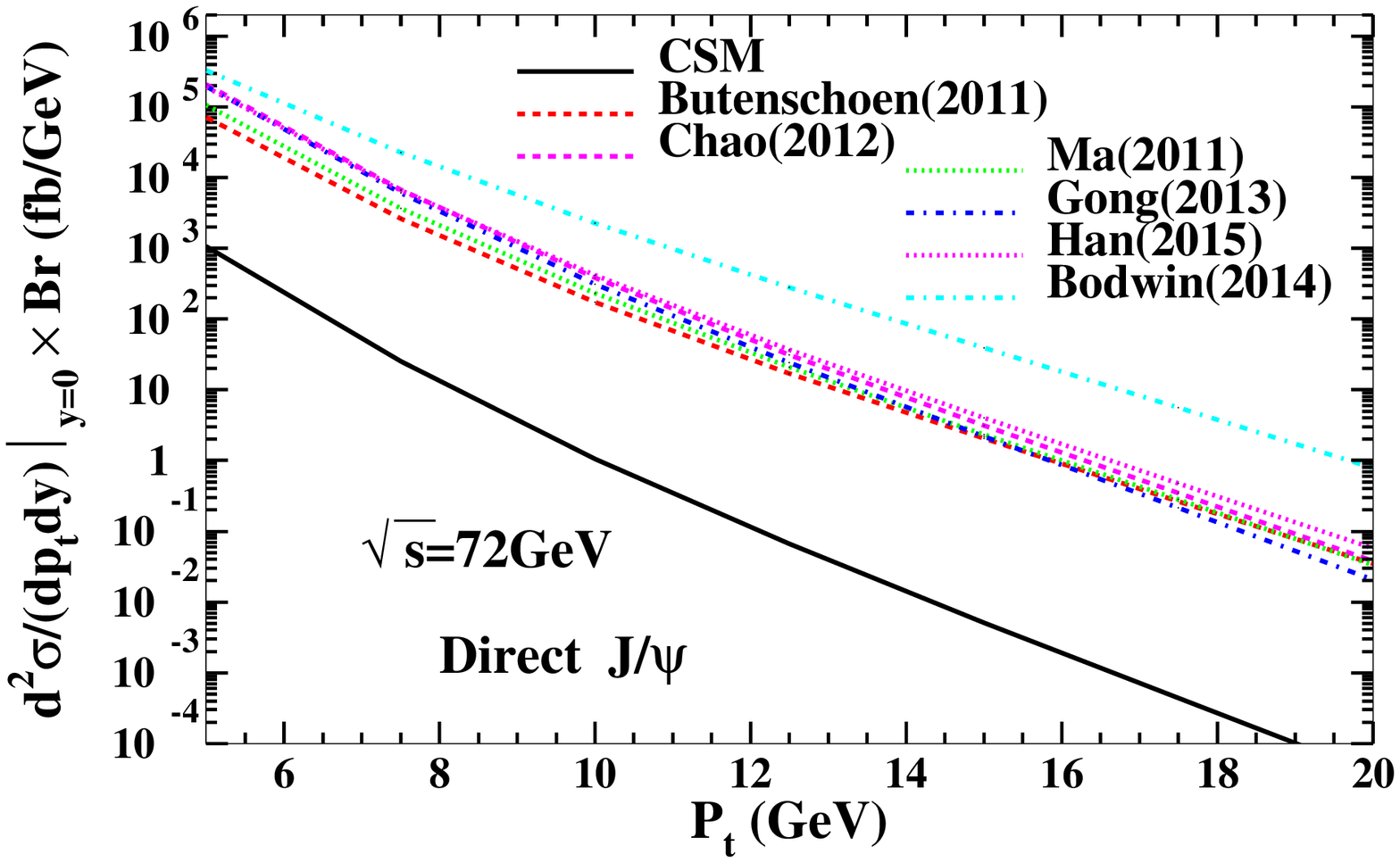}    \includegraphics[width=8cm]{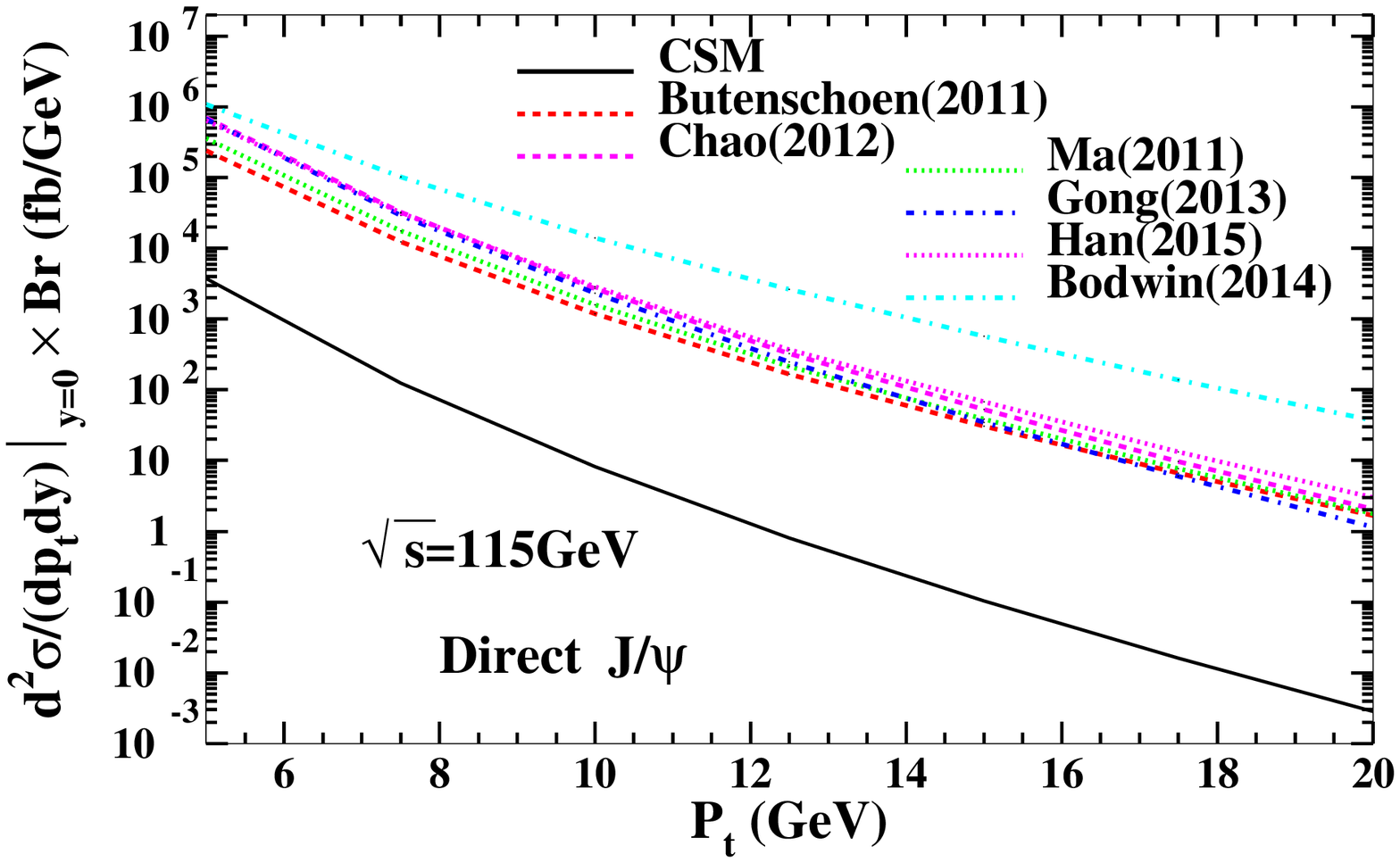} \\
  \includegraphics[width=8cm]{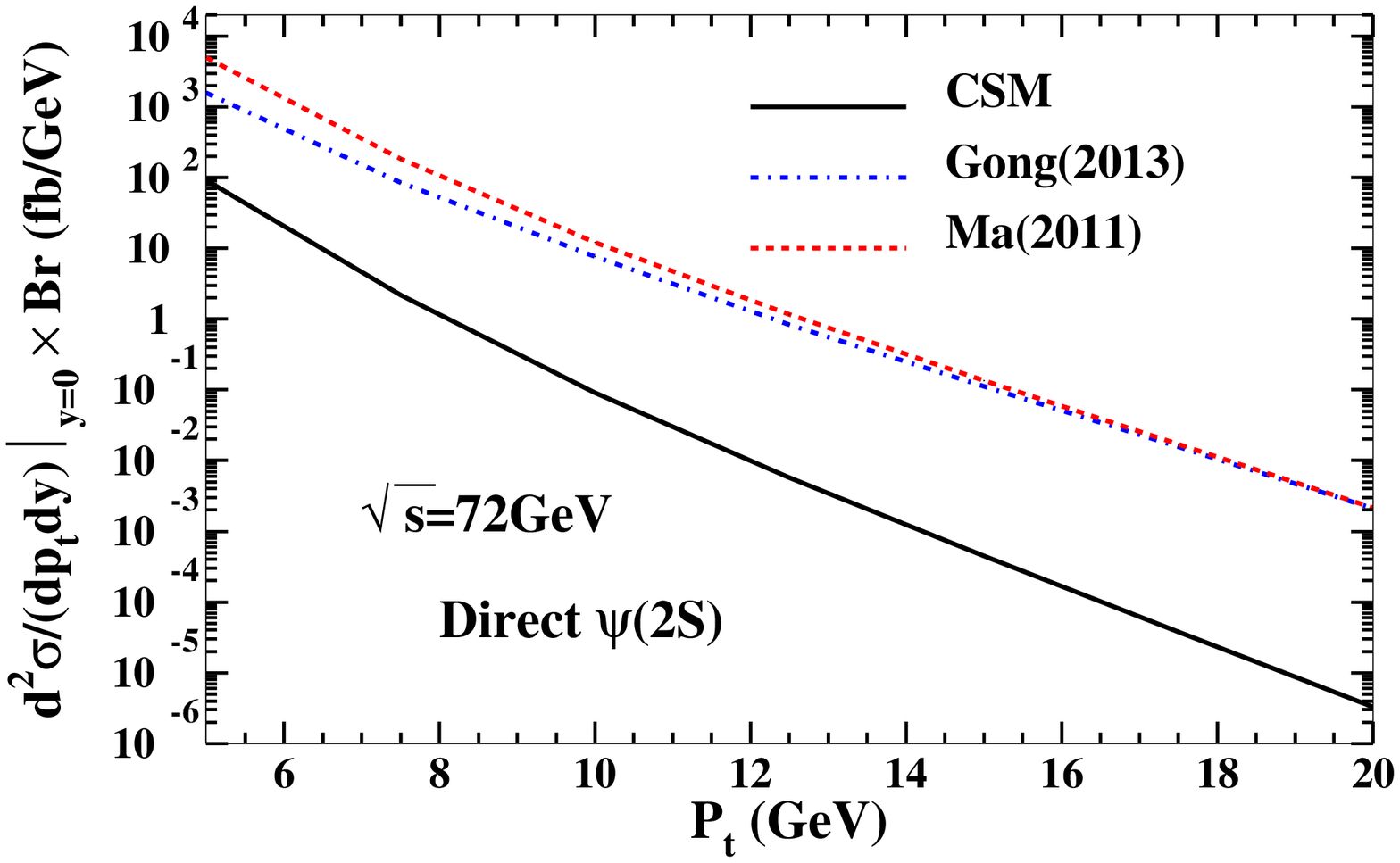}    \includegraphics[width=8cm]{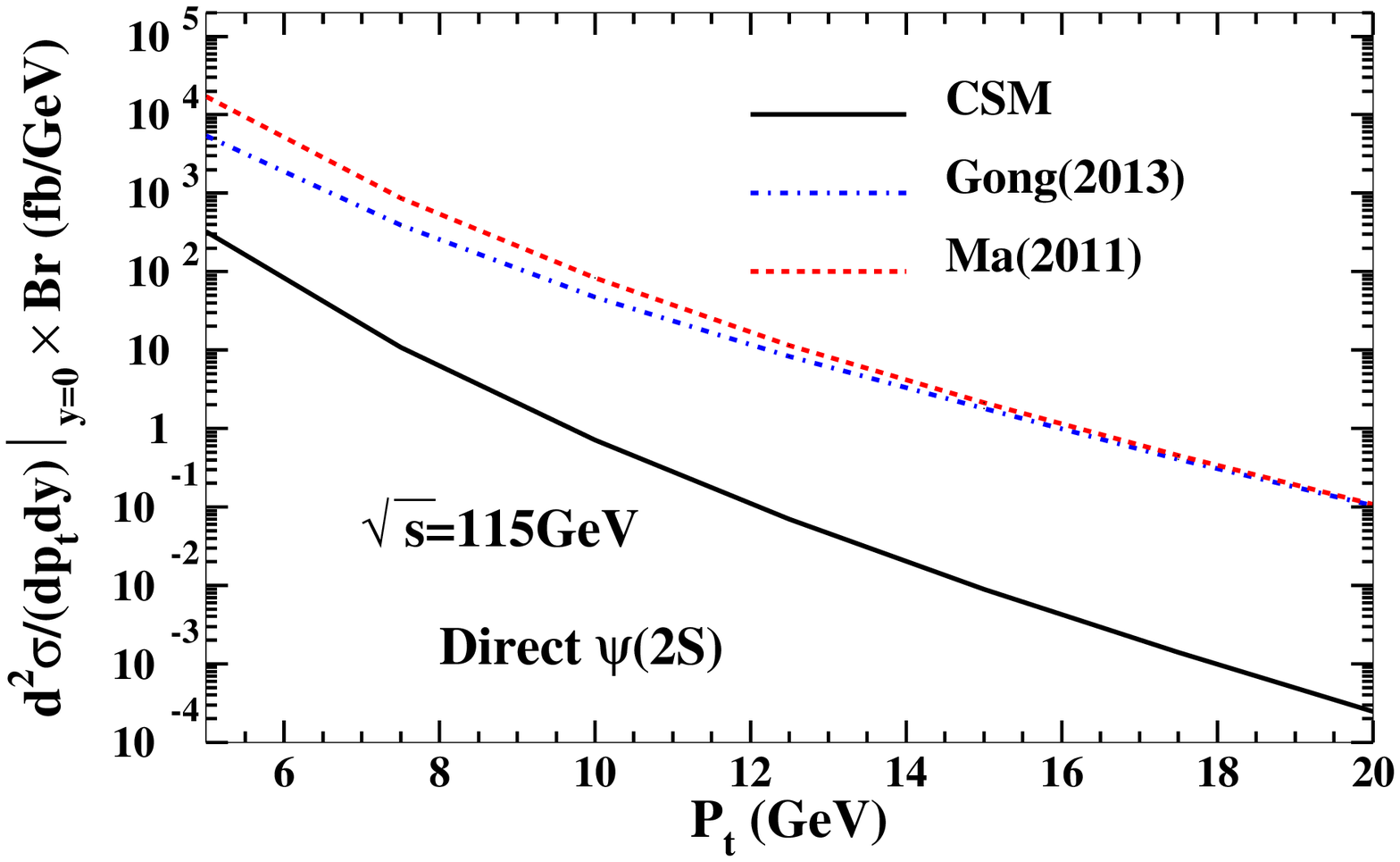} \\
  \includegraphics[width=8cm]{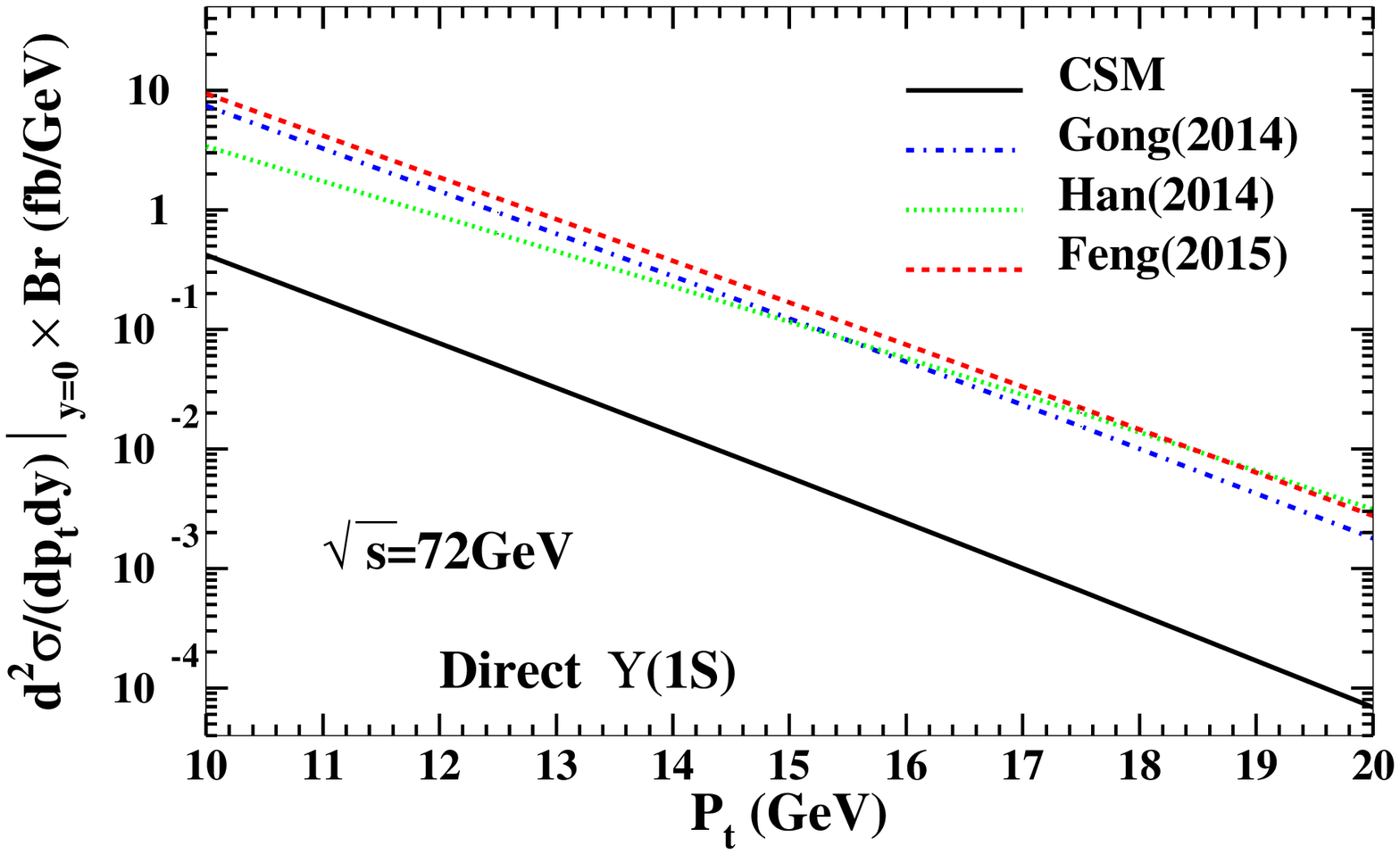}  \includegraphics[width=8cm]{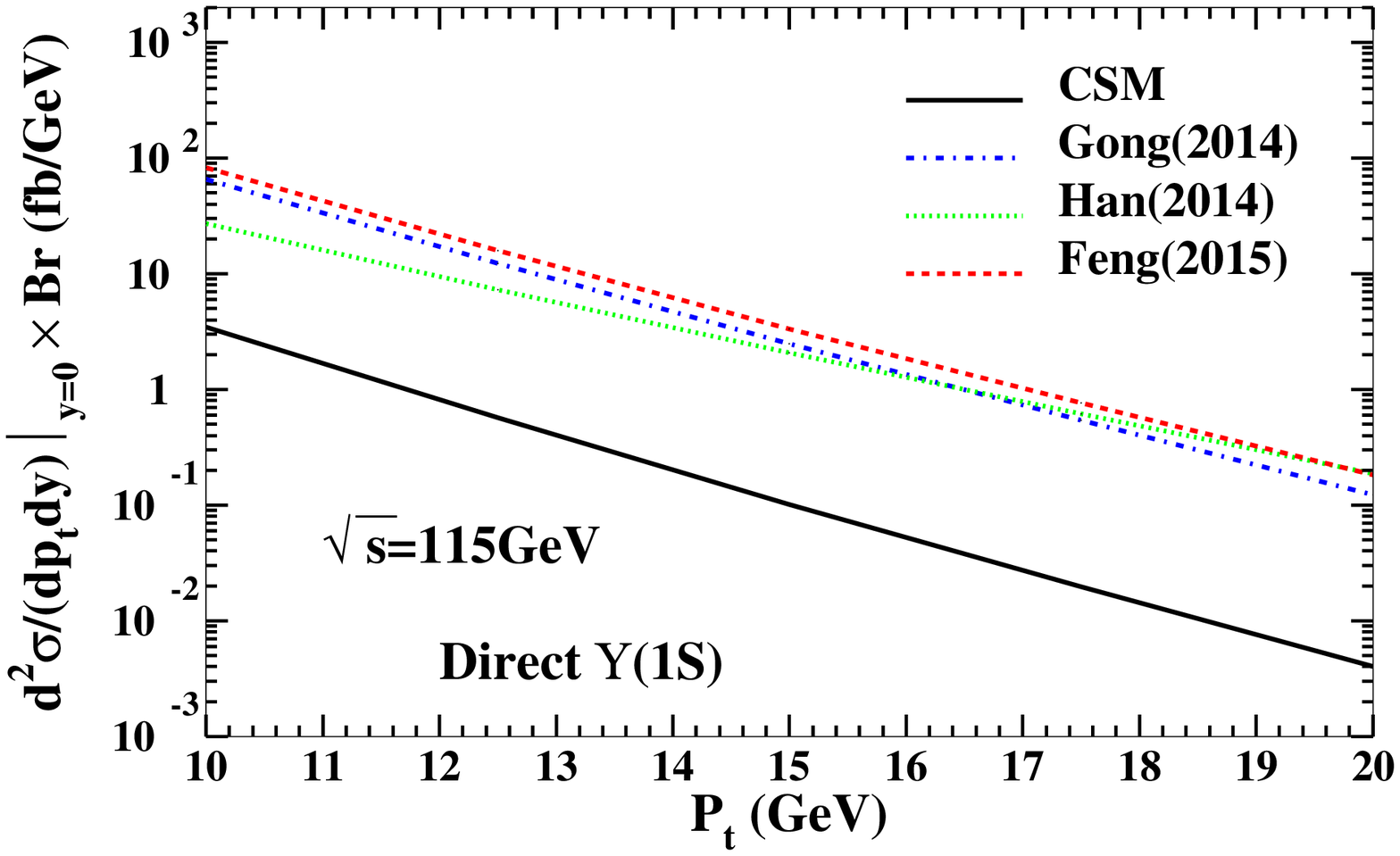} \\
  \caption{Transverse momentum distribution of differential cross section with the rapidity $y=0$ for direct $J/\psi$, $\psi(2S)$  and $\Upsilon(1S)$ hadroproduction from top to bottom, respectively. }
  \label{fig:pt-3}
\end{figure}

Now let us discuss the cross sections depend on transverse momentum $p_t$.
In Fig.~\ref{fig:pt-3}, the $p_t$ distribution of direct $J/\psi$, $\psi(2S)$ and $\Upsilon(1S)$ hadroproduction are presented.
For $J/\psi$ and $\psi(2S)$, the productions are dominated by the CO contributions, which is larger than
CSM at least one order of magnitudes that the latter one would be negligible.
The various groups of LDMEs predict $J/\psi$ and $\psi(2S)$ hadroproduction in a consistent way that the uncertainty band among them
is very narrow. Only the one from Ref.~\cite{Bodwin:2014gia} (the light blue dot-dashed line) seems to have deviated
from the uncertainty band with a larger factor 2 to 10 in $J/\psi$ case.
This may be understood by the fact that the fits in Ref.~\cite{Bodwin:2014gia} has a different SDC compared with others,
which would be the source of large uncertainty.

For $\Upsilon(1S)$, the red dashed and blue dot-dashed lines are almost parallel with little distance,
while the green dot line is obviously lower at low $p_t$ region and crosses the other ones as $p_t$ increasing.
This may explains the behavior of $d\sigma/dy$ in Fig.~\ref{fig:y}, that the low $p_t$ difference between the green
curve and the other two leads the visible distance after $p_t$ integrating.

\section{Summary}
We calculated the NLO QCD correction for direct $J/\psi$, $\psi(2S)$ and $\Upsilon(1S)$ production
at fixed-target energies.
By using the LHC beams (AFTER@LHC), we can predict the differential cross sections for the kinematics of a
fix-target experiment. We studied the cross section integrated in $p_t$ as a function of the rapidity
as well as the $p_t$ differential cross section in the central rapidity region, which are up to QCD $\alpha_s^3$
and $\alpha_s^4$ corrections, respectively. To perform a reliable prediction, various groups of NRQCD long distance
matrix elements by different fitting methods are considered as well as the uncertainties from scales and quark masses.
The results are in a consistent that the uncertainties among them is narrow.
With the typical luminosity of the fixed-target mode, which allows for
yearly luminosities as large as 20 fb$^{-1}$ for both energy,
our predictions confirm that charmonium yields can easily reach $10^9$
per year and $10^6$ for bottomonia.

\acknowledgments{We are greatful to Jean-Philippe Lansberg for his generous help in this work.}


\end{document}